\documentclass[a4paper,10pt,conference]{IEEEtran}
\pdfoutput=1
\usepackage{blindtext}
\IEEEoverridecommandlockouts                              

\title{\LARGE \bf
  Deep learning neural nets for detecting heart activity
}

\author{\IEEEauthorblockN{Joseph Horvath, Lu shien, Tommy Peng, \\ Avinash Malik}
  \IEEEauthorblockA{Electrical and Computer Engineering\\
    University of Auckland\\
    New Zealand}
  \and
  \IEEEauthorblockN{Mark Trew}
  \IEEEauthorblockA{Auckland Bioengineering Institute\\
    Auckland, New Zealand}
  \and
  \IEEEauthorblockN{Laura Bear
    \thanks{This work was supported by
      the French National Research Agency (ANR-10-IAHU04-LIRYC).}\\
    \IEEEauthorblockA{Liryc- Cardiothoracic research centre\\
      University of Bordeaux\\
      Bordeaux, France}}
}

\usepackage{graphicx}
\usepackage{float}

\usepackage[utf8]{inputenc}
\usepackage{hyperref}
\usepackage{etoolbox}
\usepackage{lipsum}
\usepackage{listings}
\usepackage{amsmath}

\usepackage{color}
\definecolor{codegreen}{rgb}{0,0.6,0}
\definecolor{codegray}{rgb}{0.5,0.5,0.5}
\definecolor{codepurple}{rgb}{0.58,0,0.82}
\definecolor{backcolour}{rgb}{0.95,0.95,0.92}
\DeclareMathAlphabet{\pazocal}{OMS}{zplm}{m}{n}

\newcommand{\listingsfont}{\ttfamily}

\lstdefinestyle{mycodestyle}{
    backgroundcolor=\color{backcolour},
    commentstyle=\color{codegreen},
    keywordstyle=\color{magenta},
    numberstyle=\scriptsize\color{codegray},
    stringstyle=\color{codepurple},
    basicstyle=\footnotesize\linespread{0.9}\listingsfont,
    breakatwhitespace=false,
    breaklines=true,
    captionpos=b,
    keepspaces=true,
    numbers=left,
    numbersep=5pt,
    showspaces=false,
    showstringspaces=false,
    showtabs=false,
    tabsize=2
}
\begin{document}

\maketitle

\begin{abstract}

The prediction of heart surface potentials using measurements from the body's surface is known as the inverse problem of electrocardiography. It is an ill-posed problem due to the multiple factors that affect the heart signal as it propagates through the body. This report details research performed into a machine learning solution to signal reconstruction as well as an analysis of optimal torso electrode positioning for prediction involving different areas of the heart. The dataset contains simultaneous measurements from a large number of body surface potential (BSP) and heart surface potential (HSP) electrodes, as well as their geometric locations, recorded from an experiment using a human model. Initially, Time Delayed Neural Nets were trained and tested across all BSP to HSP relationships resulting in a slight trend of increased reconstruction correlation with decreased separation of electrodes. However, the TDNNs had overfitted to the data and failed to predict alternate heartbeat pacings. Feed Forward Neural Nets (FFNNs) were tested in a many BSP to many HSP prediction method. Again overfitting occurred. To reduce overfitting, the number of training signals was reduced by analysing the optimal training BSPs for each HSP when using basic perceptrons. This analysis involved repeat sampling and ranking of different BSP combinations, initially, using a Monte Carlo approximation, until being replaced with a meta-heuristic which increased the yield of successful BSP combinations. Successful reconstructions across heartbeat pacings were produced using these optimal BSP combinations for 80 of the 108 HSPs, and future work exists for the testing of this method of prediction using real patient data.

\end{abstract}
\section{Introduction}

\subsection{Background}
Cardiovascular disease in New Zealand is estimated to affect 186,000 people and cause 33\%{} of deaths each year \cite{heartStats}. From a government address of the issue in 2013, it was estimated that cardiac disease had accounted for \${}501 million dollars worth of public hospital costs over the previous two years \cite{CostofHeart}. Research that could help reduce these costs is of a high importance. Irregular heartbeats can be caused by heart tissues failing to conduct their electrical signal properly. This can be the result of tissues being damaged from heart attacks. Surgeons require a diagnosis of the particular heart tissues that are damaged, before a surgery to ablate them can be performed \cite{11}.

\subsection{Current Diagnosis Method}
In order for medical professionals to diagnose damaged or improperly functioning tissues in the heart, information about the electrical activity of the heart must be measured. Currently, a common method is to perform a Cardiac Catheterization procedure. This involves the insertion of a catheter up through a leg vein and into the heart, where it can measure the electrical potentials on the surfaces of ventricles and atria. These measurements are called Electrograms or EGMs \cite{9}. EGMs represent the signal of the Heart Surface Potential (HSP) of the particular spot they were taken. Unfortunately, these catheterization procedures carry a significant risk of complications. They can cause strokes, heart attacks, and even death \cite{1}. The procedure is especially dangerous when performed on young children. The signals from the heart are conducted through the body and so can be measured on the surface of the skin in a much safer manner. These measurements are called Electrocardiograms or ECGs. ECGs represent the signal from Body Surface Potentials which are descended from the original HSPs, but are not the same. 12-lead ECG data has been used as selection criteria for Cardiac Resynchronization Therapy (CRT) using the prolonged QRS duration or the left bundle branch block morphology. However, about one third of CRT patients do not show positive responses from it \cite{2}. This is likely due to a failure to adequately locate and quantify electrical inhomogeneities from the ECG data \cite{2}. This is difficult due to many other variables adding noise to the heart signal recorded in the ECG, as well as the difficulty in relating different ECG readings to different locations in the heart. EGM data is very useful to doctors when diagnosing tissues in the heart which are not functioning correctly. If information about the EGMs was able to be extracted from ECG data, the task of diagnosing heart conditions could be made simpler. The task of reconstructing EGM signals from ECG data is known as the inverse problem of electrocardiography or Electrocardiographic imaging (ECGi) \cite{3} and is a large area of research in the mathematical and machine learning communities \cite{6}.

\subsection{The Inverse Problem}
The inverse problem of electrocardiography refers to the reconstruction of heart surface potentials (HSPs) from ECG readings. It has been mathematically proven to be ill-posed \cite{1}. This means that the problem is not uniquely determined. In other words, there exists multiple cardiac electrical signals that can give rise to the same ECG data \cite{6}. This is due to the multitude of variables that affect the propagation of the epicardial heart action potentials to the body's surface. These variables include but are not limited to, the distance, the null current across the surface of the body, and the properties of the tissues the action potentials are propagating through \cite{7}. The non-uniqueness of the problem has made it difficult for mathematical models to create accurate reconstructions. In order for more accurate predictions to be made, an alternate method must be found.

\begin{figure}[h]
  \caption{Representation of the inverse problem of electrocardiography}
  \label{inverseProblem}
  \centering
  \includegraphics[width=3in, height=3in, keepaspectratio]{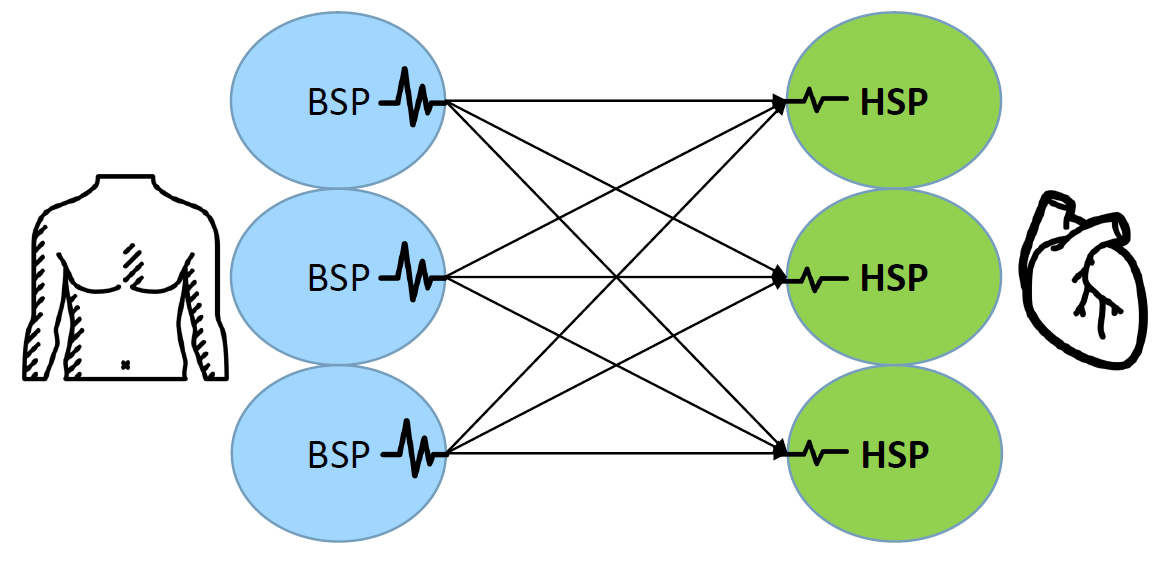}
\end{figure}

\subsection{Previous Work}
Malik, Peng, and Trew  \cite{1} details the application of a TDNN to perform ECGi on real-world recorded data from patients suffering from serous cardiac conditions. The base Artificial Neural Network structure built to predict HSPs from BSPs consists of three main components, an input layer that reads the input signals, an output layer which produces the HSP, and a hidden layer in-between. Neural net weights were chosen to minimise the mean square error using the Levenberg algorithm. The activation function used was a sigmoid function. The TDNN works based on the fact that there is high correlation between the potential now to the potential at the same position during a previous heart-beat, due to the rhythmic nature of the heart. An exhaustive search of delay window size was run using varying delay window sizes between 0 and 20. The maximum value of 20 was obtained from the correlation of BSP training time series. The optimised number of neurons in the hidden layer was determined using cross validation techniques where the neuron number was varied between 1-20 and then 40, 80, and 100 neurons. The TDNN was trained and then tested using BSP and HSP data. The BSP data was recorded from a lead attached to the electrocardiogram machine while the HSP was recorded simultaneously from the right ventricular apex. To train the TDNN, P – 1 BSP and HSP time series data sets were used. To test the TDNN, a single patient recording was used. The Pearson correlations between measured EGMs and generated EGMs were measured for both regular heart rhythms and patients with ventricular flutter. Results showed that the predicted HSPs were close to the recorded values for normal patients and that all voltage peaks were met. The correlation rates for patients with ventricular flutter were even higher with a correlation coefficient of 0.9. The average correlation across all experiments was 0.7. The best performing set up for the TDNN was with a small number of neurons and a small delay window as too large a value for these 2 parameters lead to overfitting of the TDNN during training. This study shows good potential for the use of TDNNs when performing one to one prediction of HSPs \cite{1}.

\subsection{Initial Research Goal and Hypothesis}
The main goal going into this research was to perform an analysis of the relationship between the EGM reconstruction ability of BSP electrodes and their location on the torso surface in relation to the location of the HSP that they are trying to predict. The initial hypothesis was that the closer a BSP electrode is to an HSP electrode, the higher the accuracy of the reconstruction it is able to produce will be.

\subsection{Report Structure}
This report has three main sections, One to One Reconstruction with TDNNs, Many to Many Reconstruction with FFNNs, and Analysis of Electrode Positioning. Each section describes, in regard to its own topics, the motivation for the experiments, the methodology used, and the results obtained. Lastly, there is an overarching discussion over all three sections.
\section{Dataset}
\subsection{LIRYC: University of Bordeaux}
This dataset was created and supplied to us by researchers at the heart modelling institute LIRYC at Bordeaux University. The organisations full name is: L'Institut de Rythmologie et Modélisation Cardiaque à Bordeaux. LIRYC have been researching cardiac imaging along with many other diagnosis and treatment methods for two decades, and have made major scientific progress with therapeutic applications being established across the world \cite{LIRYC}.

\subsection{Experimental Procedure}
\begin{figure}[h]
\caption{Experimental setup used by researchers at LIRYC to measure HSP and BSP signal simultaneously.}
\label{InriaSetup}
\centering
\includegraphics[width=3in, height=3in, keepaspectratio]{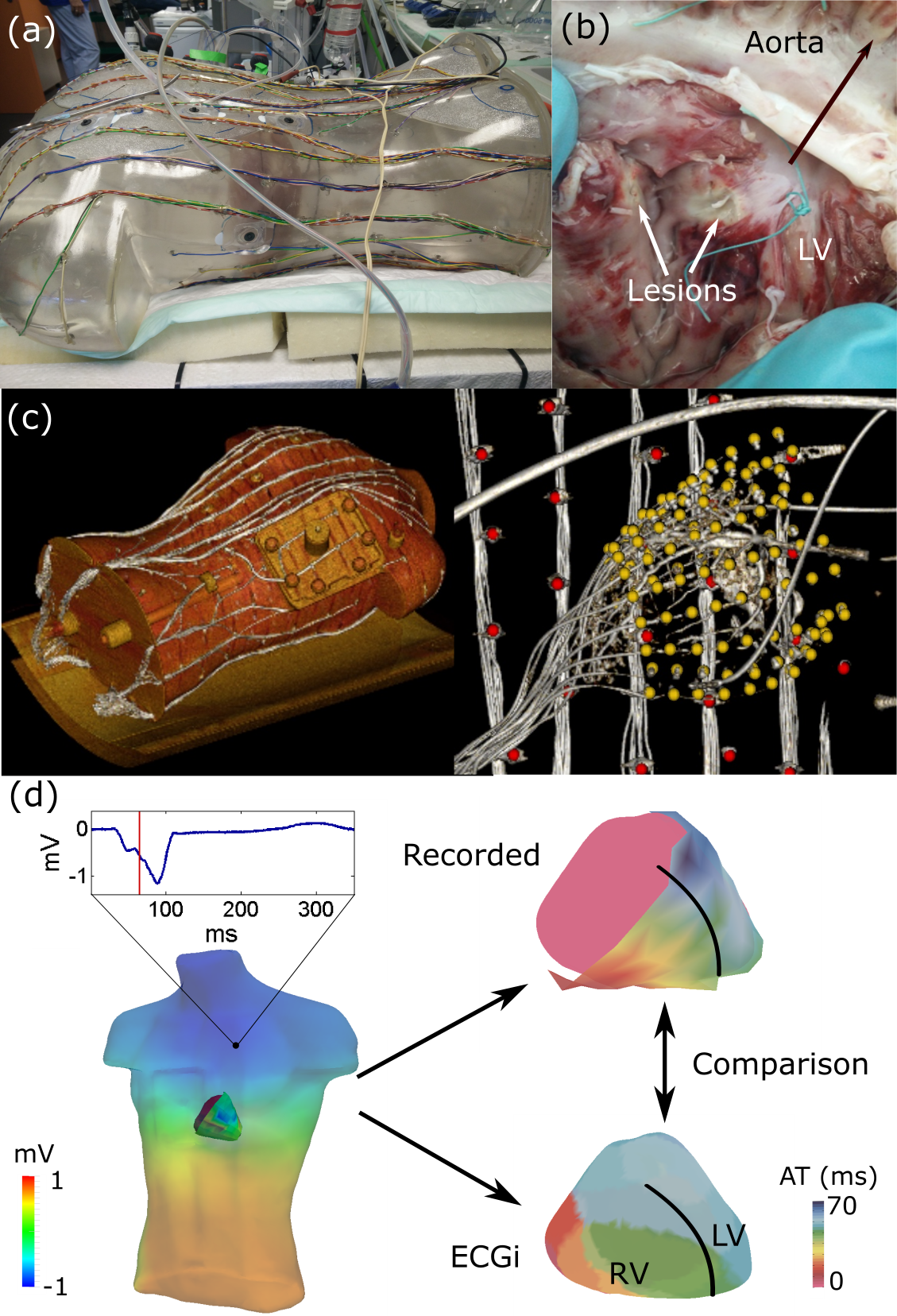}
\end{figure}

\begin{figure}[h]
\caption{Electrodes as part of the heart sock, and the LAD line orientation of the sock on the heart.}
\label{LADline}
\centering
\includegraphics[width=3in, height=3in, keepaspectratio]{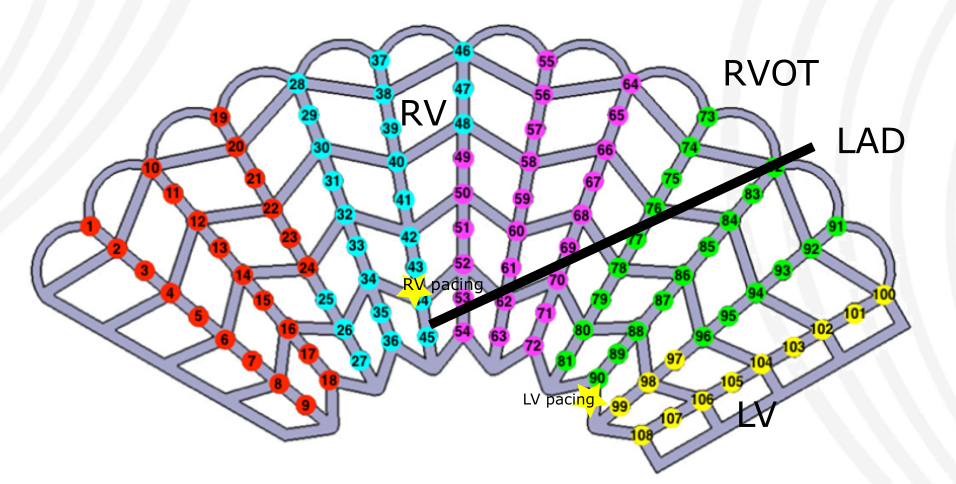}
\end{figure}

The following describes the procedure used at LIRYC to create the dataset. In order to obtain a large number of simultaneous HSP measurements, an amount that is not possible through a cardiac catheterization procedure, the model shown in Figure \ref{InriaSetup} was built to represent a human. Hearts were excised from pigs and then perfused in Langendorff mode with a 1:9 mixture of oxygenated blood and Tyrode’s solution. An epicardial sock of 108 electrodes was put over the heart and attached to the ventricles and RV apex. The sock was oriented in relation to the Left Anterior Descending Artery. The geometric vector this artery gives, shown as the ‘LAD line’ in Figure \ref{LADline}, provides information about the orientation of the epicardial sock in relation to the heart. An ablation catheter tip was placed in the LV and stitched over the bundle branch. The heart was then inserted into a human torso shaped tank of 100 percent Tyrode’s solution. The tank was covered in 128 electrodes. The heart was then able to be induced to beat with 2-5 local radio-frequency ablation procedures. The electrical potentials of all tank and heart electrodes were then able to be recorded simultaneously. The heart was stimulated in four different areas, Sinus LBBB, LV, RV, and BiV, to create different ‘pacings’. These for pacings represent the signals of the heart in different diseased states. Computed tomography was used to obtain the geometric positions of the electrodes with respect to the tank \cite{2}.

\subsection{Format of Data}
\begin{figure}[h]
\caption{The geometric locations of all HSP and BSP electrodes during experimentation.}
\label{Geometric Location}
\centering
\includegraphics[width=3in, height=3in, keepaspectratio]{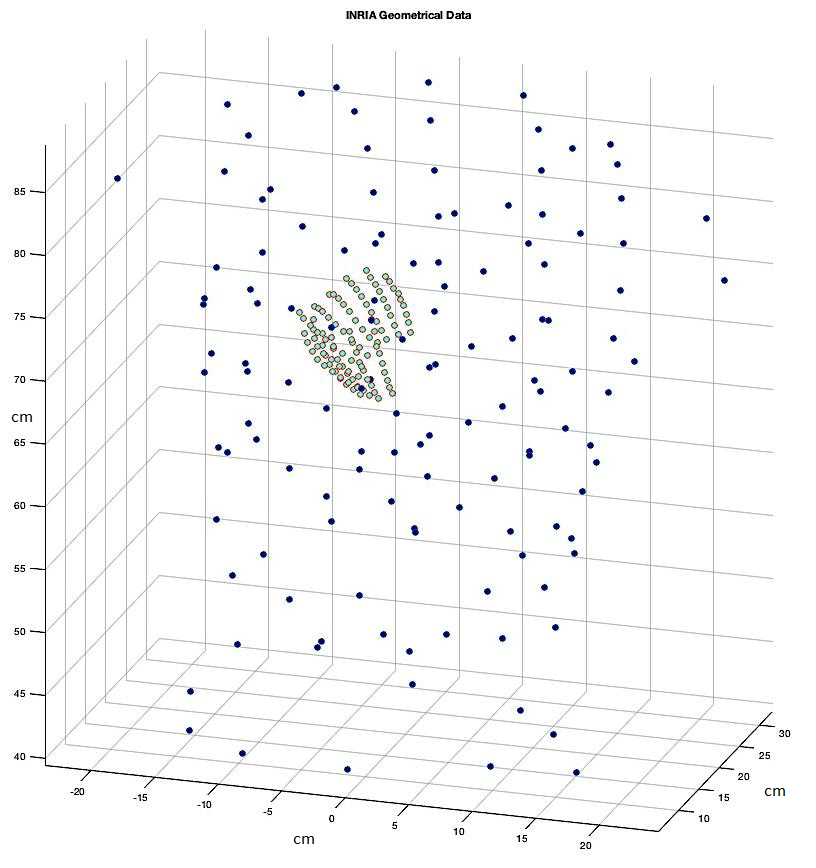}
\end{figure}

The resulting dataset is in the format of 5 MATLAB files. Four of the
files contained the electrical signal of each electrode in millivolts
over time in milliseconds. The signal data is available in many
different forms, but in these experiments only average beat and single
beat were used. Single beat contains unfiltered signals over a single
heart beat. Average beat contains a signal representing the average of
all of the beats. When using these signal sources, smoothing was applied
to them in order to remove any noise. There is also a record of which
electrodes were faulty and did not record the signal properly. The fifth
MATLAB file holds the geometric 3d coordinates of all the electrodes in
relation to each other. This allows the electrodes to be visualised as
seen in Figure~\ref{Geometric Location}.

\section{One to One Reconstruction with TDNNs}
The initial goal was to determine the optimal reconstruction for every HSP across all of the potential BSP training sets. Additionally, an examination was performed into whether there is a relationship between the locality of the BSP electrode to the HSP electrode and the accuracy of the reconstruction able to be achieved. The basis of this hypothesis was that an electrode closer to the electrode measuring the signal to be reconstructed would have less interference from interstitial ‘tissues’, or in this case fluid, and so would be closer already to the signal to be reconstructed.

\subsection{Time Delayed Neural Nets}
Time Delayed Neural Nets (TDNNs) are a type of Neural Net (NN) that operate on data that changes over time or with some other continuous or semi-continuous sequence. They work by utilising the dependence of a data point to data points prior to it in the sequence. So in the case of electrical signal data, the signal at any point in time is dependent on the signal leading up to that point. The TDNN assigns weights to a 'delay window' of inputs, and filters these inputs to help predict the next inputs. Generally, higher weights are assigned to more recent inputs as they will have a greater influence on the current signal than older inputs.

\subsection{Exploring all BSP to HSP relationships}

\subsubsection{Experimental Methodology}
The experimental procedure used when testing the HSP reconstruction ability of the TDNN with various BSPs is similar to the methodology used in Malik, Peng, and Trew’s study [1]. The tests were run using the TDNN from the Deep Learning Toolbox in MATLAB 2018a. For each BSP to HSP relationship, a semi-exhaustive search of the meta parameters was run.

\begin{table}[h]
\centering
\caption{Parameters of all BSP to HSP relationships experiment}
\label{table:paramsTDNN}
\begin{tabular}{ll}
Parameter            & Values searched through \\
Delay window size    & 5,15,30                 \\
Hidden layer neurons & 5, 10, 15               \\
Training algorithms  & "trainlm", "trainscg"   \\
Training Data        & Average beat (Smoothed), RV        \\
Testing Data         & Single beat (Smoothed), RV        
\end{tabular}
\end{table}

Each TDNN is trained using the average beat data and then tested using the single beat data. This means that the single beat being used for testing is never seen before data, although, it is incorporated into the average beat. However, this should not have much of an effect on the neural net's ability to predict this particular signal due to the numerous other beats that are incorporated in the average. In order to determine the accuracy of the reconstructions, Pearson’s Coefficient is calculated. This metric reflects the degree to which the reconstructed signal matches the measured signal. It has a maximum value of 1 corresponding to an exact replication of the signal, and a minimum value of -1 corresponding to no correlation at all.

\subsubsection{Results}
\begin{figure}[h]
\caption{A reconstructed HSP signal.}
\label{reconstructionTDNN}
\centering
\includegraphics[width=3in, height=3in, keepaspectratio]{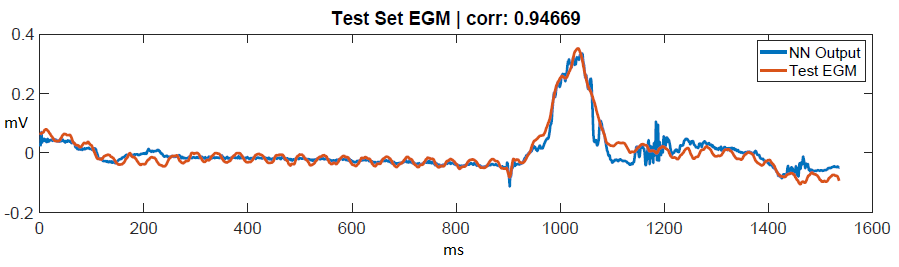}
\end{figure}

\begin{figure}[h]
\caption{BSP to HSP relationship correlation achieved using a TDNN versus the distance of the two electrodes apart.}
\label{distanceApart}
\centering
\includegraphics[width=3in, height=3in, keepaspectratio]{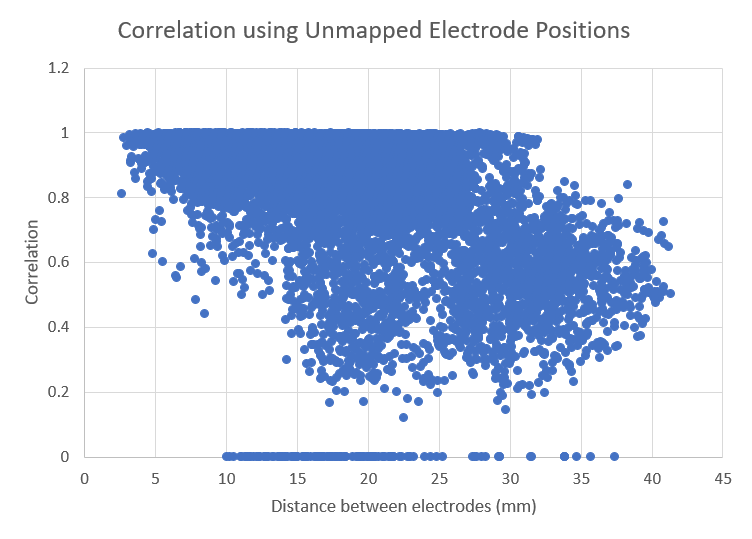}
\end{figure}

Initially, the training data and testing data both came from the RV pacing. The results from this showed that the TDNN had learned the relationships in the training data very well, with an average peak correlation achieved for all the HSP reconstructions of 0.972872. The accurate reconstruction of a single HSP can be seen in Figure \ref{reconstructionTDNN}. Additionally, the average correlation achieved across all BSP to HSP reconstructions was 0.855167, which is also a very high value. Unfortunately, the values shown here are unrealistically high and reflect the fact that the TDNN has overfitted to the training data. From Figure \ref{distanceApart}, there appears to be some kind of trend that BSP electrodes that are close to HSP electrodes are more likely to be able to predict them accurately (ignoring the faulty electrode that always found zero correlation). The trend is difficult to see, likely due to the overfitting creating vastly, high correlation reconstructions.

\subsection{Top 10 BSP to HSP Relationships}
In order to confirm the suspected overfitting of the TDNN, additional testing was performed using the top ten BSP to HSP relationships from the initial testing.

\subsubsection{Experimental Methodology}
In this experiment, as there were fewer relationships to test, the search was made more exhaustive. The trained TDNNs were tested across all the different pacings to see how they would perform reconstructing signals that are less conforming to the training data. The different pacings will have different propagation patterns to one another, while maintaining the geometry of the body. In order for the TDNN to reconstruct reliably it must learn this geometry.

\begin{table}[h]
\centering
\caption{Parameters of top 10 TDNN cross-pacing experiment}
\label{table:paramsTop10}
\begin{tabular}{ll}
Parameter            & Values searched through \\
Delay window size    & 1, 5, 10, 15, 20, 30, 40                \\
Hidden layer neurons & 5, 10, 15, 20, 40, 60, 80              \\
Training algorithms  & "traincgf", "traincgp", "trainscg"   \\
Training Data        & Average beat (Smoothed), RV        \\
Testing Data         & Single beat (Smoothed), All Pacings        
\end{tabular}
\end{table}

\subsubsection{Results}
\begin{figure}[h]
\caption{Distribution of BSP to HSP TDNN results.}
\label{TDNNgraph}
\centering
\includegraphics[width=3in, height=3in, keepaspectratio]{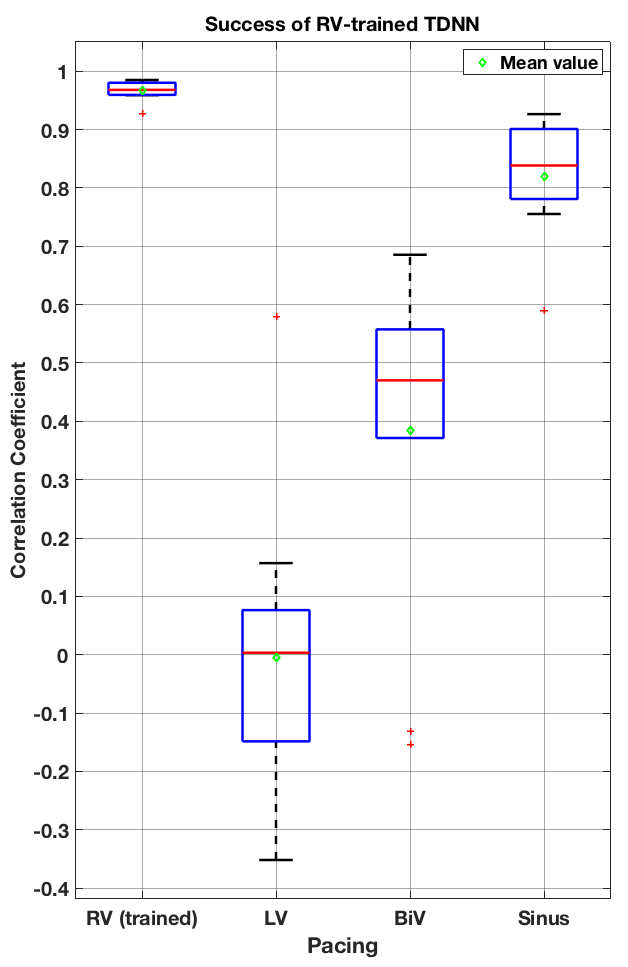}
\end{figure}

The TDNNs are shown to overfit again, giving a very good performance when tested with the RV pacing with an average correlation of 0.966 and a very small variance as shown in Figure \ref{TDNNgraph}. The problem with the overfitting is brought to light by the performance of the TDNNs when using different pacings for testing, achieving lower correlations for BiV and Sinus-LBBB, and a negative correlation for LV.

\subsection{Overfitting}
The overfitting occurring in these experiments means that the TDNNs would be unable to accurately reconstruct signals for heartbeats that are in different states of disease. This may be due to a range of factors. The fact that the TDNNs were trained using only one pacing means that they are likely missing signal propagation relationships that occur with these alternate heartbeat states. Another likely factor contributing to the TDNNs' inability to generalise is the fact that they were trained with too much data. Using such large amounts of data when training neural nets can cause them to invent relationships within data that do not actually exist. Another possibility is that in order to overcome the ill-posed problem of HSP reconstruction, more than one BSP signal is required. It seems likely that only having one perspective of a signal that can be altered throughout its propagation is not sufficient when required to reverse these changes.
\section{Many to Many Reconstruction with FFNNs}
Incorporating multiple BSPs into the reconstruction of an HSP is not readily possible using the MATLAB TDNN function which only allows the input of a single signal. In order to utilise TDNNs with this approach, a meta-classifier approach would be required to merge the outputs from multiple TDNNs. In addition, to train multiple TDNNs when predicting a single HSP is a time-consuming and CPU intensive process. As such, an approach using a different type of neural net called a Feed Forward Neural Net (FFNN) was tried.

\subsection{Feed Forward Neural Nets}
FFNNs are a very simple type of neural net consisting of perceptrons that feed forward into each other \cite{FFNNs}. Perceptrons hold matrices that filter the inputs into the outputs with varying weights. Rather than taking a time-dependent input signal, FFNNs use the scalar value of a signal at any point in time to predict the scalar value of the reconstructed signal at the same point in time. This means that multiple potentials from electrodes all around the body can be inputted to the FFNN when it is predicting a single HSP.

\subsection{Experimental Methodology}
The training time for FFNN experiments is a lot less than that of TDNN experiments. This allowed us to perform an examination of the abilities of the FFNN when trained and tested across all pacings. In order to determine parameters to use for this examination, an initial test was performed. In this test, an exhaustive search of parameters was used in the reconstruction of a single pacing to pacing experiment. The full extent of this exhaustive search is detailed in Table \ref{table:paramsFFNN}.

\begin{table}[h]
\centering
\caption{Parameters of Many BSP to Many HSP FFNN experiment}
\label{table:paramsFFNN}
\begin{tabular}{ll}
Parameter            & Values searched through \\
Hidden layer neurons & 1, 2, 3, 4, 5, 10, 15              \\
Training algorithms  & "traincgf", "traincgp", "trainscg", “trainlm”   \\
Training Data        & Average beat (Smoothed), Sinus LBBB       \\
Testing Data         & Average beat (Smoothed), LV       
\end{tabular}
\end{table}

In this experiment, it was shown that the majority of the HSP signals were reconstructed the best by FFNNs using 3 neurons in the hidden layer with ‘traincgp’ as the training algorithm. A few HSPs were predicted better with only 2 neurons in the hidden layer, however the difference in performance between these two configurations was small enough to allow the majority case of 3 neurons to be taken as the best. These ideal search parameters were then applied to every case in the examination of training to testing pacing combinations.

\subsection{Results}
\begin{figure}[h]
\caption{Cross examination of FFNN training and testing pacings.}
\label{CrossPacings}
\centering
\includegraphics[width=3in, height=3in, keepaspectratio]{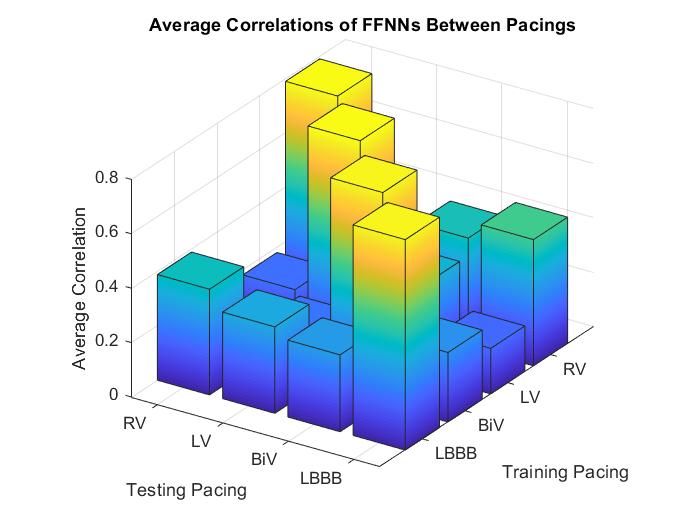}
\end{figure}

Similar to the TDNNs, FFNNs only performed to an acceptable level of correlation when tested on a pacing that was also used for training. We can see this in Figure \ref{CrossPacings} as the line of yellow bars running diagonally through the matching pacings. The best cross-pacing relationship is shown to be the RV-LBBB pairing. This pairing achieved 0.4493 average correlation across all EGMs when trained with RV and tested on LBBB and it achieved 0.3854 average correlation when trained on LBBB and tested on RV. Overfitting again is shown to be a problem when training neural nets. The amount of training data needs to be reduced. Using all 128 BSP electrodes when reconstructing HSPs was likely causing the overfitting by providing too many sources for the neural net to build data relationships from. Additionally, in a medical environment, it is somewhat impractical to fit 128 electrodes to a patient, so a solution using fewer electrodes would be more valuable.

\section{Analysis of Electrode Positioning}
In order to reduce the overfitting problem and create a more practical solution for HSP reconstruction, the number of HSP electrodes being used was to be reduced, and in doing so the optimal locations for BSP electrodes to be placed when reconstructing different HSP signals could be analysed. If an accurate reconstruction could be created using 12 well-placed electrodes on the body surface, doctors would be able to continue using their existing 12-lead ECG system in cooperation with a neural net. Additionally, aside from assisting the use of this machine learning solution, there are research opportunities for the analysis of more informative locations for doctors to place the electrodes during their existing visual analyses of ECGs.

\subsection{FFNN Parameters}
In this experiment, FFNNs, or rather perceptrons, were chosen to be the neural net used. This is because of their ability to incorporate multiple BSPs into the reconstruction of an HSP, while TDNNs are required to be incorporated into a meta-classifier. In order to determine the best combinations of BSPs to reconstruct HSPs a large amount of testing was performed. 128 possible electrodes into a 12-lead selection is calculated using 128-C-12, which equates to 2.37x10\^{}16 possible combinations of BSP electrodes. All tested BSP combinations in the analysis of electrode positioning used the same parameters shown in Table \ref{table:paramsPerceptron}.

\begin{table}[h]
\centering
\caption{Parameters of electrode positioning analyses}
\label{table:paramsPerceptron}
\begin{tabular}{ll}
Parameter            & Values searched through \\
Hidden layer neurons & 1            \\
Training algorithms  & “trainlm”   \\
Training Data        & Average beat (Smoothed), Sinus LBBB       \\
Testing Data         & Average beat (Smoothed), LV       
\end{tabular}
\end{table}

\subsection{Perceptrons}
Using a single neuron in the hidden layer of an FFNN, as seen in table \ref{table:paramsPerceptron}, means that the FFNN is now just a simple perceptron. A perceptron is essentially a matrix of weights. These weights are applied to the different inputs, filtering them to different degrees into each of the outputs. Replacing FFNNs with perceptrons made sense for multiple reasons. Experimentation had shown that, in many cases, perceptrons could achieve comparably high correlations to our 3 neuron FFNNs, although the correlations were not quite as high. However in these experiments, achieving a high correlation is lower priority than determining the best BSP electrode locations, as the optimal BSPs can be used to train a new FFNN with different parameters once found. Additionally, with fewer neurons, there is a lower chance that overfitting will occur. Using the more basic form of signal reconstruction that perceptrons apply, means that there is less analysis of the relationships between different BSP signals. Rather, the signals are being directly tested for their similarity to the HSP signal requiring reconstruction. BSPs that are already most similar to an HSP are likely the ones that are most useful for reconstruction using machine learning, or for visual examination of ECGs by medical professionals. Finally, the fewer neurons used in an FFNN, the faster it is to train. The speed of training is very important in allowing the sampling of enough BSP combinations for the results to be meaningful.

\subsection{Monte Carlo Approximation}
Monte Carlo analyses are used to produce approximations of a true function, through repeated random sampling. In our case, we will be attempting to approximate the ‘usefulness’ function of the BSPs for each HSP.

\subsubsection{Experimental Methodology}
For each HSP, 500 repeat samples were performed.  In each, combinations of 12 BSPs were created by sampling from the total 128 BSPs without replacement. The perceptrons were then trained using these 12 BSPs and then tested with the other pacing. The BSPs used in these tests were then tallied and grouped by the correlation of the reconstruction they were able to create in their combination. The groupings are in 0.05 increments and combinations were only recorded if their performance was above 0 correlation. In order to analyse these experiments as a whole, a ‘usefulness’ metric was created for BSPs. The method of determining usefulness in our data is based on the frequency of BSPs in combinations that achieved high correlations. Its calculation is outlined more thoroughly in Figure \ref{usefulnessCalc}.

\begin{figure}[h]
\input{Matlab/Usefulness.tex}
\caption{MATLAB code showing the calculation of BSP usefulness across all HSPs.}
\label{usefulnessCalc}
\end{figure}

\subsubsection{Results}
\begin{figure}[h]
\caption{3D graphic of BSP usefulness to BSP location on the torso.}
\label{3dUseful}
\centering
\includegraphics[width=3in, height=3in, keepaspectratio]{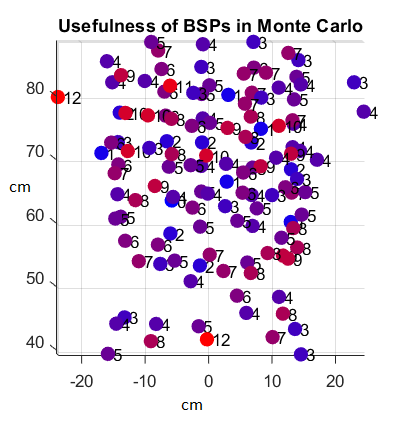}
\end{figure}

\begin{figure}[h]
\caption{2D unrolling of the torso electrodes with BSP usefulness. The electrodes were unrolled such that the left and right edges are the centre of the back.}
\label{2dUseful}
\centering
\includegraphics[width=3in, height=3in, keepaspectratio]{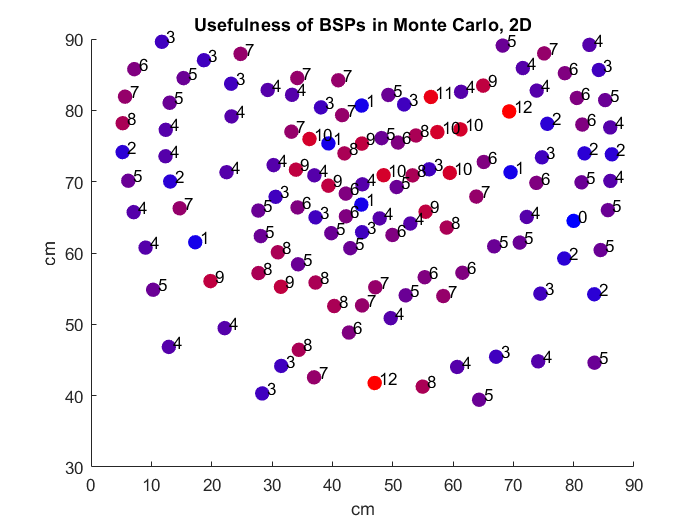}
\end{figure}

From Figures \ref{3dUseful} and \ref{2dUseful}, we can see that some BSPs were more ‘useful’ than others. Largely, it appears that BSP electrodes located on the front and centre of the chest appear to be more ‘useful’ than those located towards the shoulders, back, or lower down on the torso. This further affirms the hypothesis that locality improves a BSPs reconstruction ability, as the electrodes on the front and centre of the chest are closer to the majority of the HSP electrodes. However, there is one electrode on the stomach with a very high usefulness score of 12.

The problem using this Monte Carlo approach is that the proportion of combinations that achieve high correlations is very low. Only 1.8\%{} of BSP usages were part of combinations that achieved above a 0.8 correlation. This means that the number of BSP combinations used to determine BSP usefulness is quite small and likely is not a reliable indicator. Additionally, it means that more than 98\%{} of our experiments are not useful and a large proportion of the runtime is going to waste. In order to create more reliable results, a greater number of useful combinations is required to be generated.

\subsection{Meta-Heuristic Approach}
In order to generate more ‘useful’ combinations of BSPs in fewer tests, a switch was made from Monte Carlo approximation to a meta-heuristic approach. This method uses a learning function to guide random BSP combinations towards higher correlations using a neighbourhood search for BSPs. This allows more BSP combination to eventually achieve above a correlation threshold after sufficient swapping of BSPs.

\subsubsection{Experimental Methodology}
\begin{figure}[h]
\caption{Process of the Meta-Heuristic altering the combination of BSPs to improve the reconstruction correlation.}
\label{metaProcess}
\centering
\includegraphics[width=3in, height=3in, keepaspectratio]{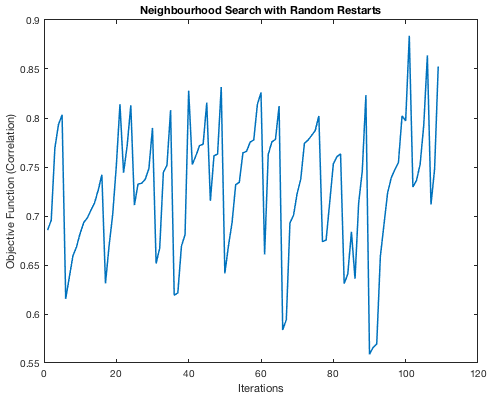}
\end{figure}

Experiments were run with combination sizes, 2, 6, 12, and 24. Random combinations of BSPs were sampled without replacement from the total 128. In each experiment, for each HSP, 500 random restarts were performed. Each random restart sees a random BSP combination created then tested. If the correlation achieved is less than the correlation threshold of 0.8, BSPs in the combination are switched out using a neighbourhood search. The meta-heuristic will continue replacing BSPs in the combination until either the correlation threshold is met, or the neighbourhood search runs out of neighbours. The number of neighbours is defined as double the size of the combination size. In Figure \ref{metaProcess}, we can see the meta-heuristic achieving the correlation threshold as the higher peaks. After each of these peaks there is a trough which corresponds to the random restart.

\subsubsection{Results}

\begin{figure}[h]
\caption{The distribution of Meta-Heuristic produced BSP-combination correlations achieved.}
\label{metaResults}
\centering
\includegraphics[width=3in, height=3in, keepaspectratio]{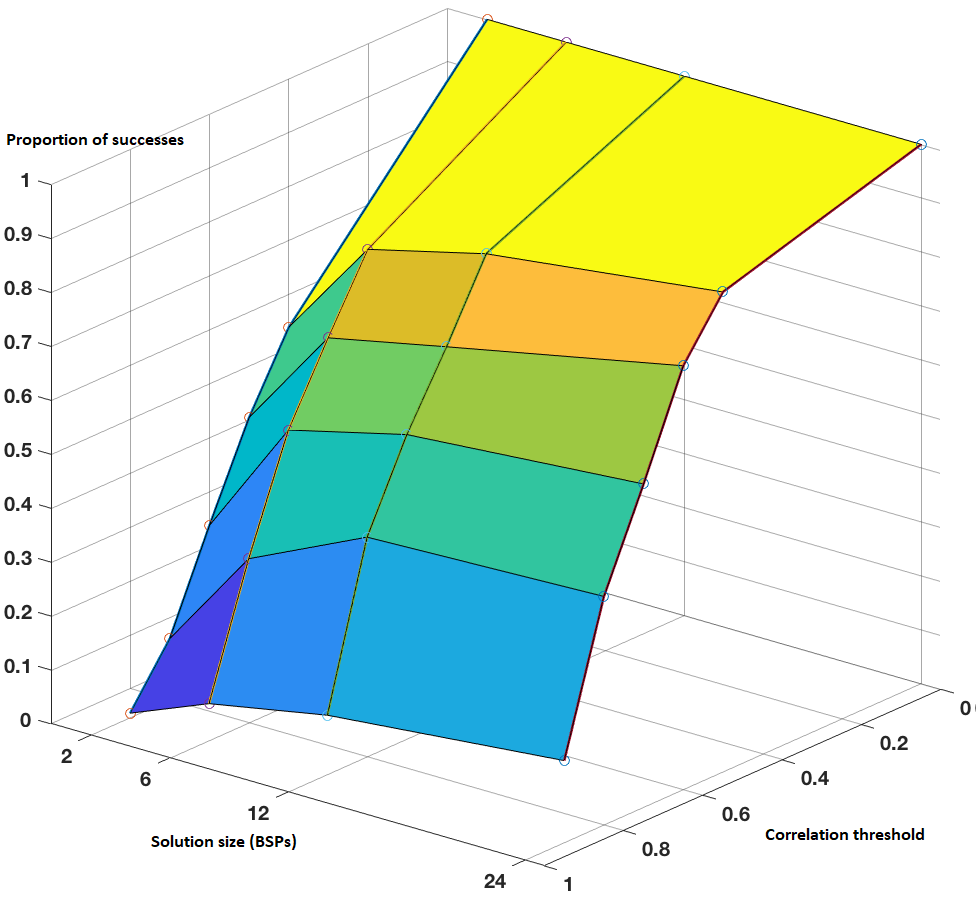}
\end{figure}

Across all the experiments, 40.67\%{} of 12-BSP combinations created by the meta-heuristic achieved above the correlation threshold of 0.8. This is a big improvement from the 1.8\%{} result of the Monte Carlo approximation. In Figure \ref{metaResults}, there is a distribution of correlations achieved by BSP-combinations across the different combination sizes. As more correlations are incorporated, the proportion of included combinations gets closer to one. There is a visible trend across the whole graph, that as the solution size increases, the proportion of combinations that achieved a certain correlation increases. This suggests that the optimal combination size across all HSPs will lie somewhere between 24 and all 128 BSPs. However, it should be less than the total 128 based on the previous experiment where FFNNs performed poorly across pacings when using all the BSPs at once. The optimal combination size can also fluctuate across HSPs.

 In order to find the optimal combinations of BSPs for each EGM, they must be looked at one by one. The result was that most of the best solutions were in the 24 BSP combination size. 28 of the HSPs had no solutions that achieved above a 0.8 correlation. This could be as a result of these locations being very difficult to predict. Alternatively, there could have been issues with the electrode connection to the heart surface during experimentation. This would result in the target signal being distorted.

\begin{figure}[h]
\caption{Mapping BSP 'usefulness' to the torso tank for HSP 77. The BSPs are painted as triangles on the torso surface with blue colour representing low ‘usefulness’, and yellow representing high usefulness.}
\label{colourTorso}
\centering
\includegraphics[width=3in, height=3in, keepaspectratio]{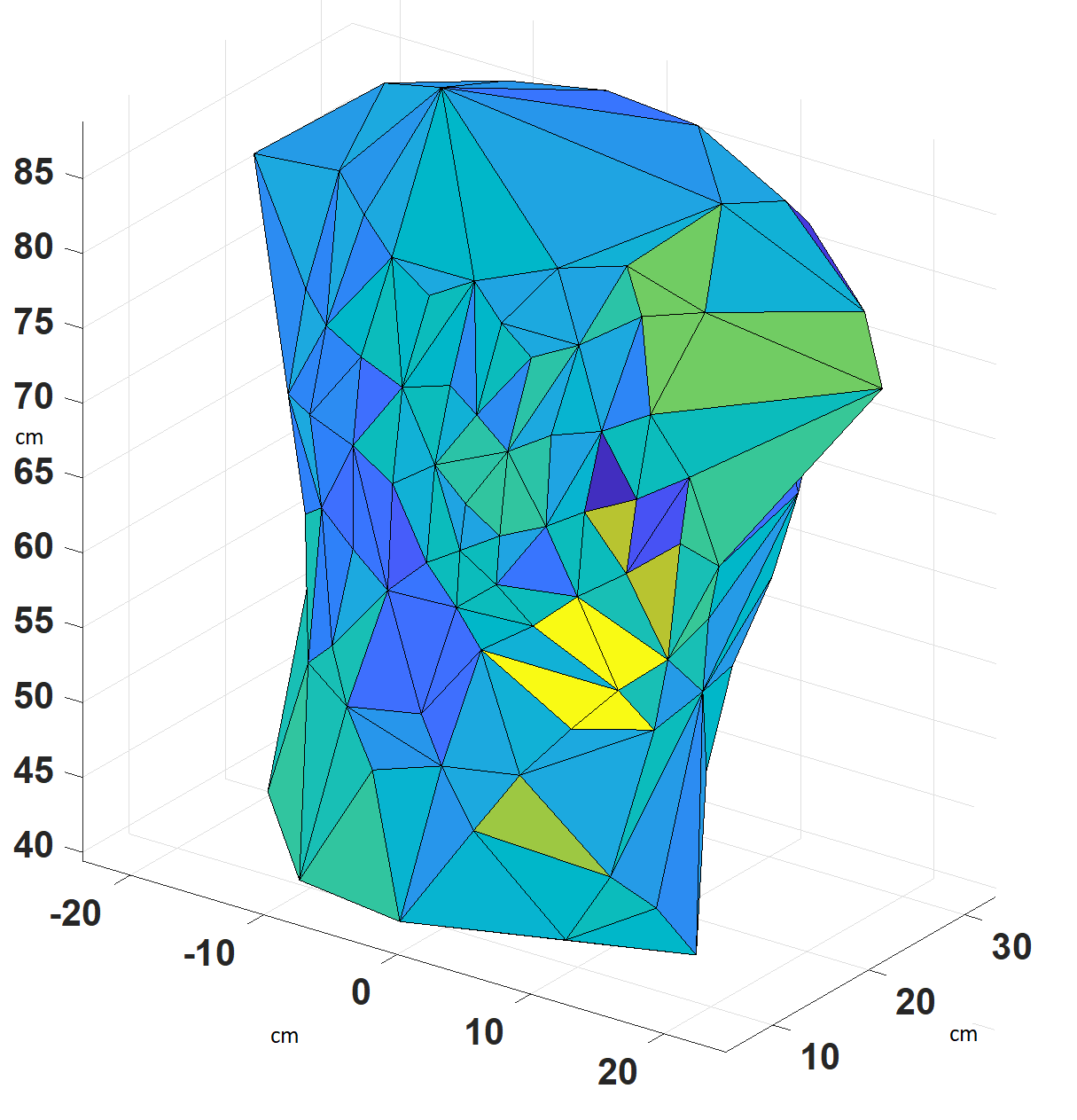}
\end{figure}

\begin{figure}[h]
\caption{Ranking of BSP 'usefulness' for HSP 77.}
\label{77ranking}
\centering
\includegraphics[width=3in, height=3in, keepaspectratio]{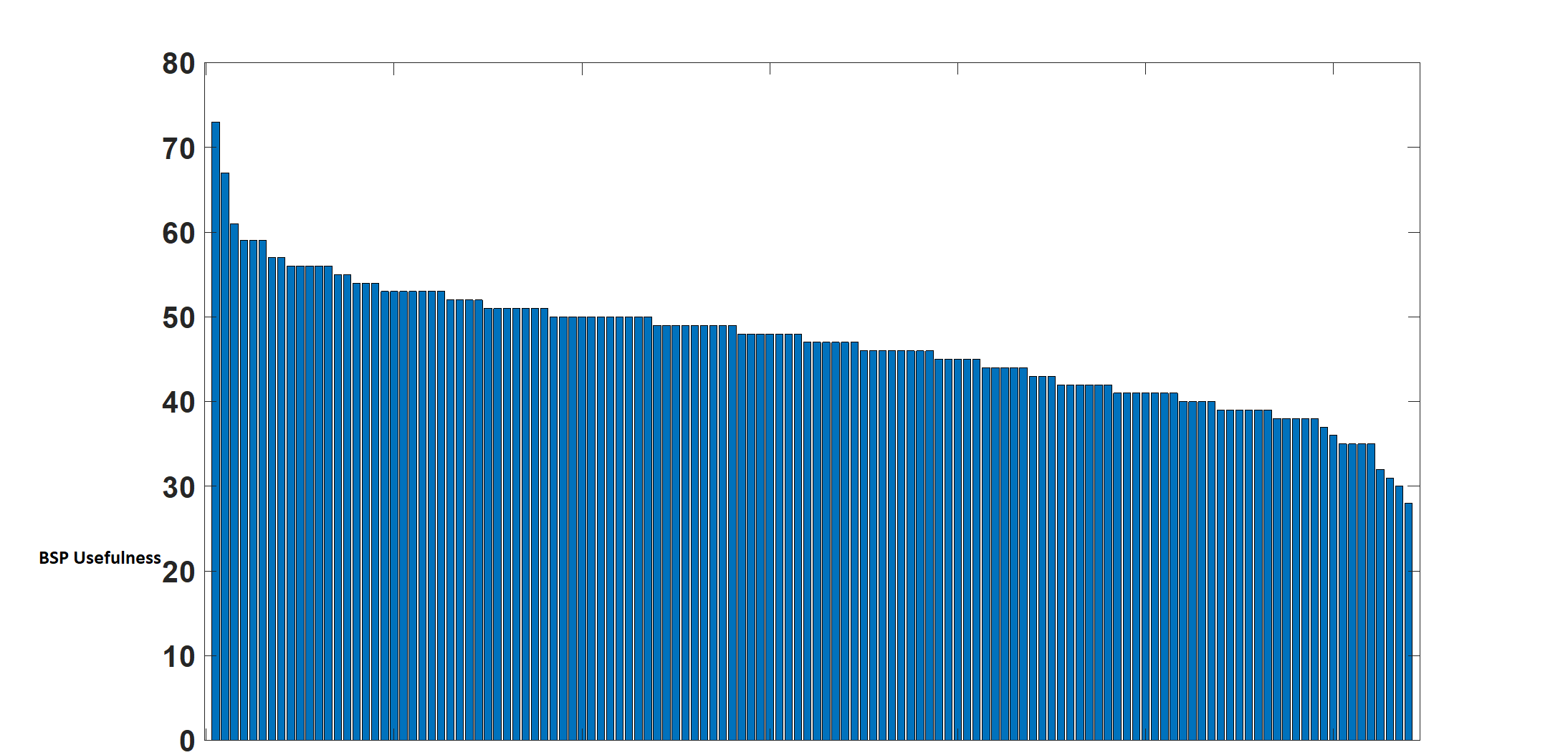}
\end{figure}

\begin{figure}[h]
\caption{The reconstructed signal of the best BSP combination for HSP 77.}
\label{77Reconstruction}
\centering
\includegraphics[width=3in, height=3in, keepaspectratio]{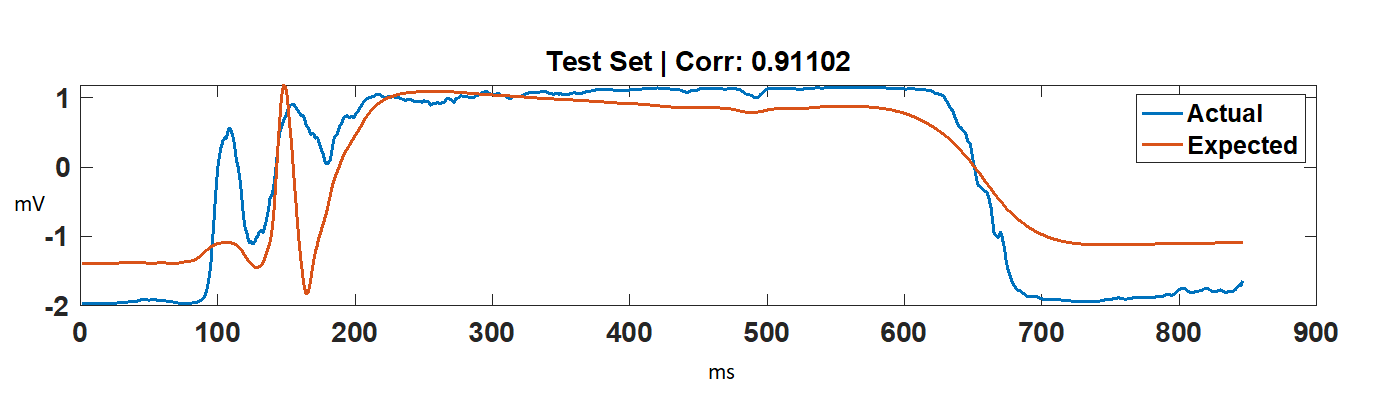}
\end{figure}

The optimal combination of each HSP was determined by ranking the ‘usefulness’ of all the BSPs and then picking the highest scoring off the top. It is not possible to show the results achieved for all 108 HSPs here, so instead the results for HSP number 77 are shown. This HSP is one that was able to be most accurately reconstructed. In Figure \ref{77ranking}, the ranking of the BSP electrodes for this HSP can be seen. Many of the electrodes achieved an average usefulness score while a few near the top left achieved significantly higher scores. These rankings can also be viewed geometrically in Figure \ref{colourTorso}. From this we can see that six of the most useful BSPs were all located near each other. This further confirms the locality hypothesis. The optimal solution for this HSP was a 12-BSP combination. The top 12 BSP electrodes from the ranking are, 121, 11, 79, 59, 90, 16, 29, 72, 32, 54, 123, and 13. This combination achieved a correlation of 0.91102. This is quite a high correlation especially across pacings, and the reconstructed signal, shown in Figure \ref{77Reconstruction}, is quite similar to the original. However, there are differences between them. The first peak in blue measured signal is not accurately matched by the reconstruction. Even though this error does not create a large drop in the Pearson’s Correlation metric, it may be of great consequence to the ability of doctors to use the reconstruction to diagnose the tissues.

Lastly, the meta-heuristic search combined with BSP ranking  used in this experiment is a powerful method of input-data selection and could have applications in other uses of machine learning.

\begin{figure}[h]
\caption{The location of heart electrode number 77 on the heart sock within the torso.}
\label{77Location}
\centering
\includegraphics[width=3in, height=3in, keepaspectratio]{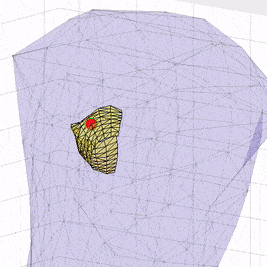}
\end{figure}

\section{Projection Mapping}
If the machine learning solution for HSP reconstruction used in this research were to be applied in practice for medical diagnosis, methods of normalisation would need to be applied across patients. Two factors that can change between patients that could affect the performance of the trained neural nets are the size of the patient’s torso and the orientation of the patient’s heart. In order to account for these variables, a process of projection mapping has been developed. Projection mapping converts the 3D geometric point locations of the electrodes in the data set to a 2D format. This allows the relationships to be found between BSP and HSP electrodes, independent of the torso size and heart orientation values of the dataset model.

\subsection{Process of Projection}

\begin{figure}[h]
\caption{The Heart sock fitted with an ellipsoid.}
\label{ellipsoid}
\centering
\includegraphics[width=3in, height=3in, keepaspectratio]{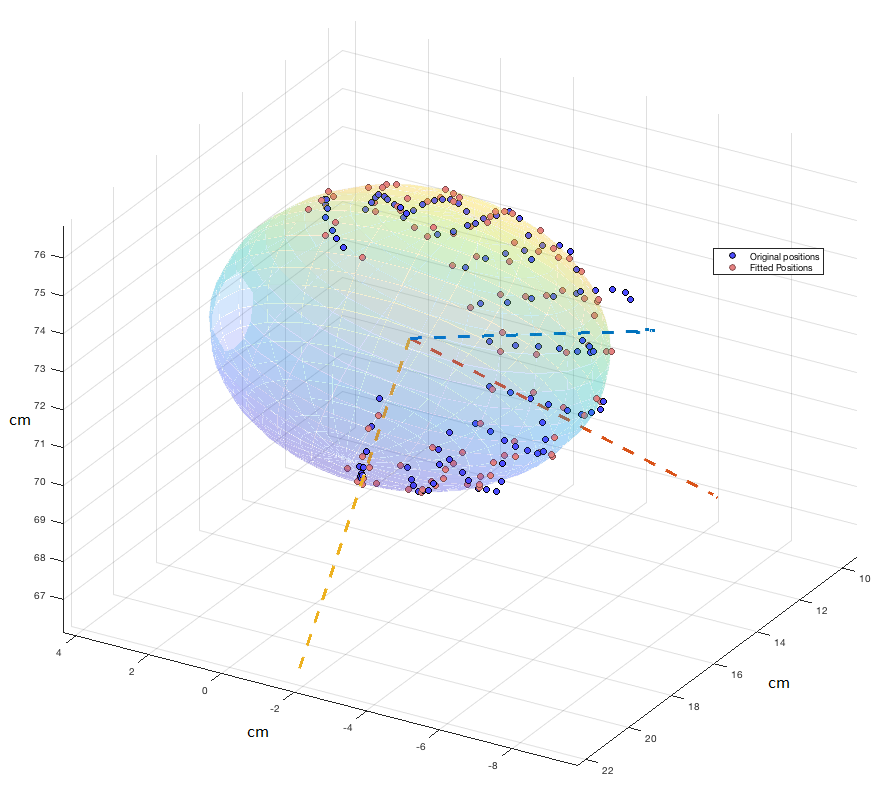}
\end{figure}

\begin{figure}[h]
\caption{The torso tank fitted with a cylinder.}
\label{Cylinder}
\centering
\includegraphics[width=3in, height=3in, keepaspectratio]{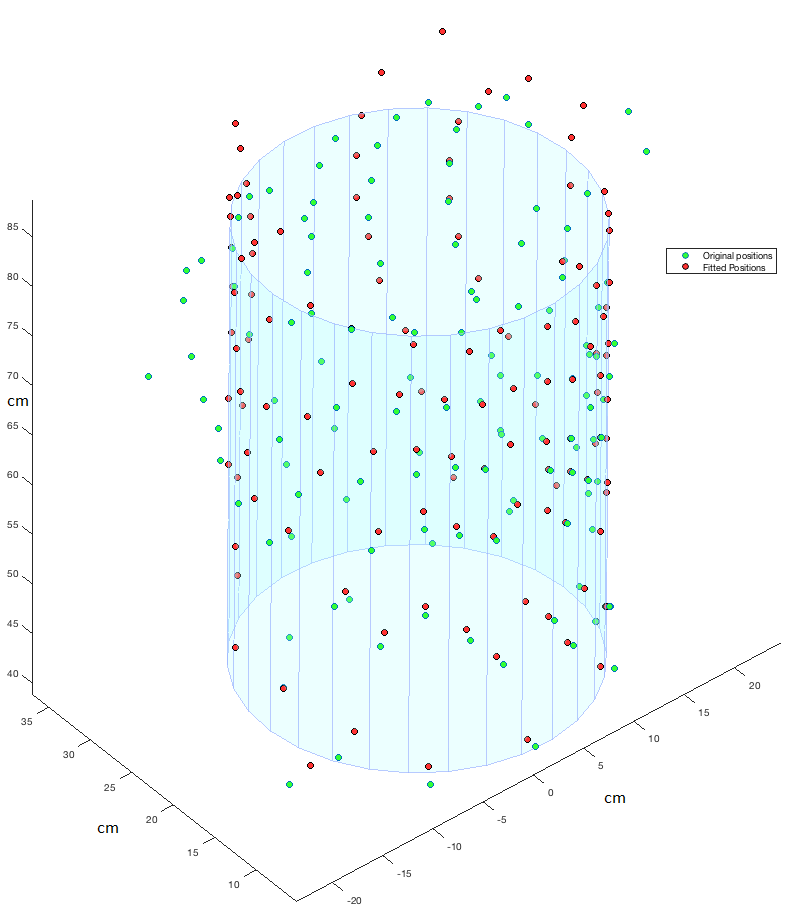}
\end{figure}

Firstly, the heart sock and torso tank geometries were simplified. An ellipsoid was fitted to the heart sock by minimising the distance between the ellipsoid surface and the heart sock electrode points (Figure \ref{ellipsoid}). A cylinder was fitted to the torso tank points in a similar manner, minimising the distance between torso tank electrode points and the cylinder surface (Figure \ref{Cylinder}).
  
\begin{figure}[h]
\caption{Projecting the heart electrodes onto the cylinder.}
\label{Projection}
\centering
\includegraphics[width=3in, height=3in, keepaspectratio]{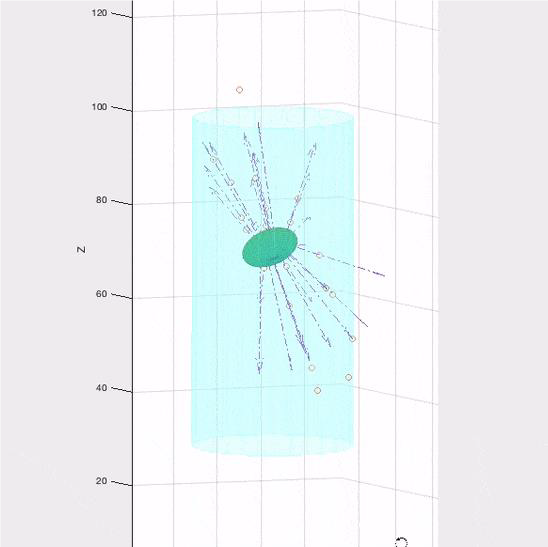}
\end{figure}

The electrode points were then projected onto this cylinder. The torso tank points were projected by simply selecting the closest point on the cylinder to each electrode. The heart sock points were projected in a different manner. Vectors were drawn through each sock point, running perpendicular to the surface of the ellipsoid. These vectors were then traced to their point of intersection with the cylinder surface. This is where the projection points were set. Heart sock projection onto the cylinder's surface can be visualised in Figure \ref{Projection}

\begin{figure}[h]
\caption{Final projection result, a 2D unrolled cylinder with projected electrode points. BSP electrodes: Green, HSP electrodes: Purple.}
\label{unrolled}
\centering
\includegraphics[width=3in, height=3in, keepaspectratio]{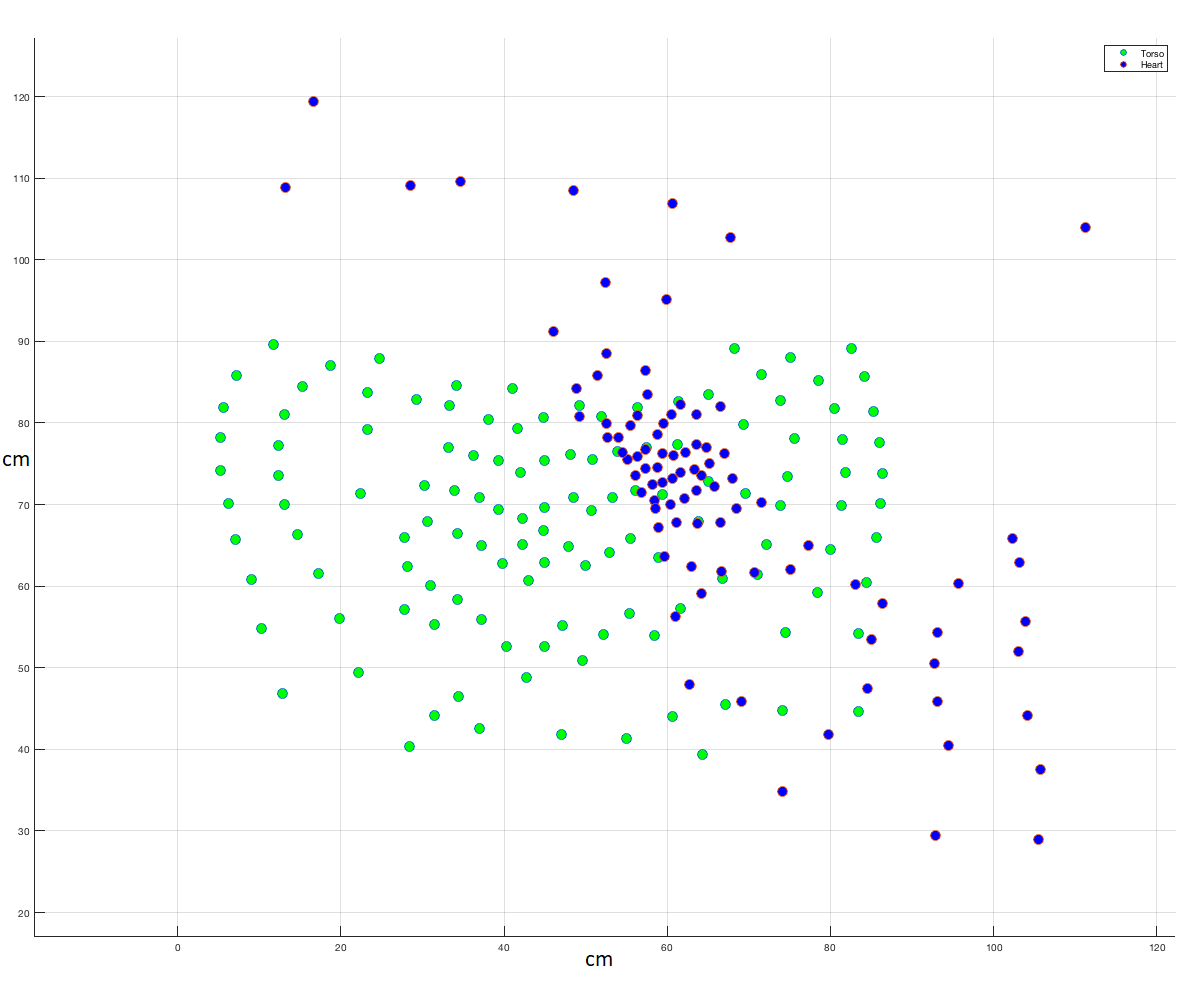}
\end{figure}

The cylinder with all the mapped points was then cut along its surface following the vector of the LAD line, and then unrolled to a flat surface which is shown in Figure \ref{unrolled}.

\subsection{Theoretical Application of Projection Normalisation}
The application of this normalisation is yet to be tested but could occur as follows. From this study, information about which BSP electrode positions give accurate reconstructions for each HSP is known. These electrodes can be located in the 2D projection. Using a CT scan, information about the orientation of the patient's heart as well as their torso size can be obtained. The process of projection could then be reversed using the measurements from the patient. The 2D projection could be rolled into a cylinder, oriented using the patient’s LAD line, and with a diameter that matched the patient's torso size. This would give the locations on the patient’s torso where the cardiographer can place their electrodes in order to gain signal information for the optimal reconstruction of  the HSP signal using the pre-trained neural net.
\section{Discussion}
The initial hypothesis that BSP to HSP locality improves reconstruction ability appears to be affirmed by many of these results. However, in some cases, using alternate perspectives of the signal from different areas on the torso, such as from the stomach area, also gave positive results. Using smaller combinations of BSPs when training neural nets appears to have reduced the overfitting problem and have been used to created high accuracy reconstructions of many of the HSP signals. There is still much analysis of this data required in order to confirm these theories.

\subsection{Threats to Validity}
There are many threats to the validity of these results. All the data used came from experiments using a model of a human patient. While the functionality of a porcine heart is quite similar to that of a human, the torso tank it was placed in is much simpler than a real human torso. The lack of any internal structures within the torso tank means that the propagation of the heart signals is much less obstructed. This would likely increase the strength of the relationship between a BSP's reconstruction ability and the locality of it to the HSP being reconstructed. The fact that some HSPs could not be reconstructed at all is unusual. There could be a possibility that the connection of certain electrodes to the heart is not as sound as the other more reconstructable HSPs. This raises the question as to which HSP signals may have been disrupted in this way. If they were, it would affect which BSPs performed better when reconstructing them. The methods of normalisation described in the projection mapping research have not been tested. Questions could be raised as to the effectiveness of the normalisation process. While the electrode positions are adjusted geometrically, there is no guarantee that the relationship between the BSP and HSP electrodes will be the same due to the many variables that can change between patient anatomies.

\subsection{Future Work}
The next step in this research would be to test the performance of the trained neural nets using real patient data. This could use EGM data from patients who require the procedure to be performed already. It would be interesting to test the optimal BSP electrode placements, along with the normalisation methods, on these patients to see if the neural nets can reconstruct the EGMs in a practical setting. There is also the potential for the use of TDNNs in a meta-classifier when using a multi-BSP perspective. TDNNs are a good fit for signal reconstruction. Combining their effectiveness at one BSP to one HSP reconstruction with a many BSP approach could have great potential in achieving high correlations.

\section{Conclusions}
The analysis of BSP electrode location in relation to its ability to create high correlation reconstructions has affirmed, if not confirmed, the initial hypothesis of electrode locality. It has also resulted in the determination of the optimal BSP combinations for each HSP. For 80 of these 108 HSPs, the optimal combination was able to reconstruct the recorded signal with at least 0.8 correlation when inputted into the neural net. The meta-heuristic ranking method used for finding these useful BSPs has been shown to be effective and holds potential for use in other training data selection problems. The combinations used for these reconstructions are of a magnitude 2-24 electrodes, which is a practical amount for cardiographers who are currently used to a 12-lead approach. Future work could involve the testing of these alternate body surface electrode layouts with real patient data, both in the form of EGM reconstruction using neural nets, or providing an alternate ECG format for doctors to visually diagnose from.
\section{Acknowledgements}
I would like to thank my project partner Lu Shien Lee for his hard work performing this research with me. Also, I would like to thank Laura R. Bear and the other researchers at LIRYC, for providing us with such an interesting dataset. Finally, a big thanks to Avinash Malik, Mark Trew, and Tommy Peng, for their guidance and teachings throughout this project, which were extremely helpful to us in this work.


\begin{thebibliography}{10}

\bibitem{LIRYC}
Cardiac imaging at liryc. retrieved from
  https://www.ihu-liryc.fr/en/patient-care/our-clinical-teams/r/cardiac-imaging/.

\bibitem{2}
Laura~R. Bear, Peter~R. Huntjens, Richard Walton, Olivier Bernus, Ruben
  Coronel, and Remi Dubois.
\newblock Cardiac electrical dyssynchrony is accurately detected by noninvasive
  electrocardiographic imaging.
\newblock {\em Heart Rhythm}, 2018.
\newblock ID: Heart Rhythm, 15(7): 1058-1069, 2018.

\bibitem{FFNNs}
George Bebis and Michael Georgiopoulos.
\newblock Feed-forward neural networks.
\newblock {\em IEEE Potentials}, 13(4):27--31, 1994.

\bibitem{CostofHeart}
National~Health Committee.
\newblock Strategic overview for cardiovascular disease in new zealand.
\newblock December 2013.

\bibitem{heartStats}
Heart Foundation.
\newblock General heart statistics in new zealand. retrieved from
  https://www.heartfoundation.org.nz/statistics.

\bibitem{7}
S.~Giffard-Roisin, T.~Jackson, L.~Fovargue, J.~Lee, H.~Delingette, R.~Razavi,
  N.~Ayache, and M.~Sermesant.
\newblock Noninvasive personalization of a cardiac electrophysiology model from
  body surface potential mapping.
\newblock {\em IEEE Transactions on Biomedical Engineering}, 64(9):2206--2218,
  2017.

\bibitem{1}
Avinash Malik, Tommy Peng, and Mark Trew.
\newblock A machine learning approach to reconstruction of heart surface
  potentials from body surface potentials.
\newblock {\em arXiv preprint arXiv:1802.02240}, 2018.

\bibitem{11}
F.~Porée, A.~Kachenoura, G.~Carrault, R.~D. Molin, P.~Mabo, and A.~I.
  Hernandez.
\newblock Surface electrocardiogram reconstruction from intracardiac
  electrograms using a dynamic time delay artificial neural network.
\newblock {\em IEEE Transactions on Biomedical Engineering}, 60(1):106--114,
  2013.

\bibitem{6}
A.~J. Pullan, L.~K. Cheng, M.~P. Nash, C.~P. Bradley, and D.~J. Paterson.
\newblock Noninvasive electrical imaging of the heart: Theory and model
  development.
\newblock {\em Annals of Biomedical Engineering}, 29(10):817--836, 2001.

\bibitem{3}
Charulatha Ramanathan, Ping Jia, Raja Ghanem, Daniela Calvetti, and Yoram Rudy.
\newblock Noninvasive electrocardiographic imaging (ecgi): Application of the
  generalized minimal residual (gmres) method.
\newblock {\em Annals of Biomedical Engineering}, 31(8):981--994, 2003.

\bibitem{9}
N.~Zemzemi, S.~Labarthe, R.~D. Dubois, and Y.~Coudière.
\newblock From body surface potential to activation maps on the atria: A
  machine learning technique.
\newblock In {\em 2012 Computing in Cardiology Conference}, pages 125--128,
  2012.

\end{thebibliography}

\end{document}